# Ontology-based Classification and Analysis of non-emergency Smart-city Events

Monika Rani*, Sanchit Alekh, Aditya Bhardwaj, Abhinav Gupta and O. P. Vyas
Department of Information Technology,
Indian Institute of Information Technology,
Allahabad, India
*monikarani1988@gmail.com

*Abstract*—Several challenges are faced by citizens of urban centers while dealing with day-to-day events, and the absence of a centralized reporting mechanism makes event-reporting and redressal a daunting task. With the push on information technology to adapt to the needs of smart-cities and integrate urban civic services, the use of Open311 architecture presents an interesting solution. In this paper, we present a novel approach that uses an existing Open311 ontology to classify and report non-emergency city-events, as well as to guide the citizen to the points of redressal. The use of linked open data and the semantic model serves to provide contextual meaning and make vast amounts of content hyper-connected and easily-searchable. Such a one-size-fits-all model also ensures reusability and effective visualization and analysis of data across several cities. By integrating urban services across various civic bodies, the proposed approach provides a single endpoint to the citizen, which is imperative for smooth functioning of smart cities.

## I. INTRODUCTION

Around the world, cities aspire to provide their residents better access to civic services and a superior quality of life. However, as cities grow, developing applications for urban areas that could improve the management of urban flows and allow for real time responses to challenges becomes a pressing need. In several Indian cities and urban centers, the individual entities and different government bodies are sparsely connected that leads to a poor flow of resources and information residents encounter in their daily lives, a variety of events like disruptions in services and utilities, inconvenience caused by elements in the environment, or need for information and updates from the civic authorities. Due to the dissociated nature of the bodies, they often find themselves in a difficult situation to handle such events, e.g., if someone wants to report about stagnant water on the streets that have persisted for more than a few days, they are usually unaware about whom to contact. In such a situation, citizens have to approach multiple authorities before reaching the concerned ones.

Although it is difficult to precisely define a 'smart city', Deakin and Al Waer list four factors [1] that contribute to the definition of a smart city: (i) The application of a wide range of electronic and digital technologies to communities and cities. (ii) The use of ICT to transform life and working environments within the region. (iii) The embedding of such ICTs in government systems. (iv)The territorialisation of practices that brings ICTs and people together to enhance the innovation and knowledge that they offer. Therefore, ICT plays a vital role in information management and analysis in the event and data-driven smart cities. At the same time, the expedited growth of population and access to technology in cities necessitates the generation of huge amounts of data. But city-data originating from heterogeneous sources cannot be effectively integrated or analyzed unless they share common vocabulary and semantics [2]. A semantic data model of a city is, therefore, the most feasible solution as it not only provides cohesive and integrated data access across city's data sources, but also leads to reusability and can be used for replicating the same across other cities as well. An important breakthrough in this endeavor has been the introduction of the Open311 open standard for civic issue tracking. As a technical standard, it's a protocol that software systems can implement to create interoperable systems. The Open311 standard has been successfully implemented in various cities around the world such as Helsinki, Chicago, Toronto, San Francisco among many others. An exhaustive list can be found on the Open 311[1] official website.

'311' is the contact number and name, used primarily in North America, of organizations that help to provide responses to non-emergency requests and reports submitted by the citizens. An ontology enables the integration of 311 data from a myriad of cities with consistent semantics [3]. In our paper, we have used an Open311-ontology[2] created at the University of Toronto by Nalchigar and Fox to annotate and map non-emergency city events to a semantic framework, and provide residents a single endpoint to report, query and reach the concerned authorities according to their complaint. A complete and authoritative ontology simplifies the development of applications that require integrated access to city data sources and enables solution reuse as we move from one city to the next [4]. Moreover, in order to make the entire process extremely user-centric and usable even for the technologically non-inclined user, we have made it possible for the citizen / user to register his query in natural language instead of SPARQL [5]. Apart from interconnecting government services, the annotated information gained from the mechanism would make it possible to be able to understand the current rapidly changing business environment, derive insights and make data-driven decisions.

---

[1] http://www.open311.org/
[2] http://ontology.eil.utoronto.ca/open311.owl

## II. LITERATURE REVIEW OF RELATED WORK

Semantic Web [6] forms the basis of this paper, with a goal to form a global web of machine readable data. It provides a common framework that allows data to be shared and reused across applications, enterprise, and community boundaries. With the help of Semantic web, data can be retrieved using the general web architectures, e.g. Uniform Resource Identifier (URI). Since the application of semantic web includes efficient data integration, resource discovery and classification, it presents a functional way of integrating information and resources from a variety of different departments in a city.

Semantic web also publishes languages that are specifically designed for data, like Resource Description Framework (RDF) [7], Web Ontology Language (OWL) [8], and Extensible Markup Language (XML) [9]. With these technologies, it's possible to describe arbitrary subjects and aspects of a city, such as traffic, water supply, landscaping, garbage, etc., using a standard, open, expressive, schema flexible and interoperable architecture.

An ontology defines a set of representational primitives with which to model a domain of knowledge or discourse. The representational primitives are typically classes (or sets), attributes (or properties), and relationships (or relations among class members) [10]. A key advantage of an ontology is that it allows abstraction and independence from the implementation and low level data models. This advantage of an ontology [11] facilitates a straightforward integration of homogeneous databases and enables interoperability between diverse systems. The edge of integration and interoperability makes an ontology a rational solution for representing the complex nature of Indian cities, where exists a sundry of subjects and sparse information on them. This proves to be crucial as the smart-cities project must be replicated in a large number of Indian cities for it to be beneficial for everyone. Gainful information can be extracted from the ontology [12] with ease, after different departments and their scope, relations and information are clear and evidently laid out by the ontology, with the use of a querying language.

The use of semantic web and the event-driven paradigm for modeling smart cities was proposed by Cretu, who stated that events are more than just data, so that meaningful information is represented which can be used by people or software agents for any kind of specified actions [13]. They presented a concept and architecture for an event-driven, smart-city and described the technologies that can be used to implement such a system. There were several efforts to integrate city data from heterogeneous sources, however, the introduction of the ISO 37120: Sustainable development of communities -- Indicators for city services and quality of life, gave the cities standard indicators on which their performance could be compared [14]. Moving further in this direction, Uceda-Sosa et al. By IBM Research, as a part of IBM's smart cities initiative, developed a comprehensive ontology for representation of several cities-related facts, knowledge, organization, services, flow of events and messages and key performance indicators. The model was an engineered attempt towards consumable, industry-strength semantic models for cyber-physical systems [4]. In the PolisGnosis project, Fox tried to overcome the weaknesses of the SCRIBE ontology by automating the longitudinal and transversal analysis of city indicators based on the ISO 37120 standard [2]. Thereafter, Nalchigar and Fox developed an Open311 ontology to represent the Open311 service calls from various cities in a common semantic framework, to allow for better reusability and analysis of data [3]. Vlacheas et al., in their work, identified the main issues with the use of the Internet of Things (IoT) in smart cities, and instead, proposed a cognitive management framework for IoT, in which they represent dynamically changing real-world objects in a virtualized environment [15]. Also, Consoli et al. proposed an urban fault reporting and management system designed and developed for the municipality of Catania. The system was based on a data model that describes user fault alerts and is linked to a complete e-government data model developed within the PRISMA project. Although domain specific ontologies have been used for integrating city-services, a comprehensive literature survey reveals that no significant work has been done to map city events across various government and civic departments in the Indian scenario. Moreover, the previous ontologies are not compatible with the Open311 standard, which is increasingly being used in most urban agglomerations [16]. Our paper tries to provide a solution to these issues.

## III. METHODOLOGY

In this paper, we have used the Open311 ontology to map non-emergency city events and direct the citizens towards points of redressal. For running experiments and SPARQL queries on the Open311 ontology in this paper, we have created instances of locations, events, messages and service requests for a specific test group, i.e. our university- Indian Institute of Information Technology, Allahabad (IIIT-Allahabad).

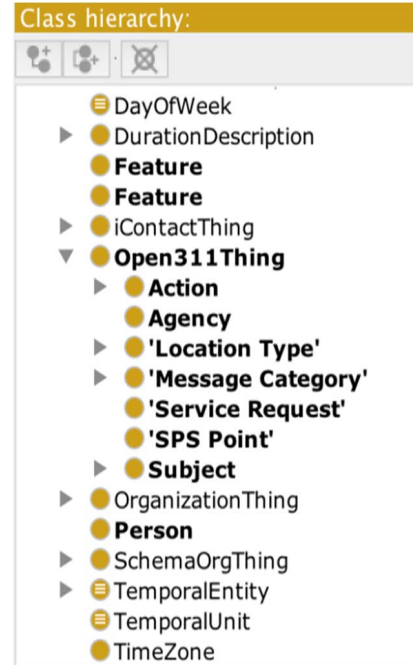

Fig. 1. Class Hierarchy in the Open311 ontology

The scope of the presented approach is limited to 8 types of events, namely Tree, Grass, Insect, Waste, Street Light, Smoking, Noise and Internet. Further, there are 6 locations that have been considered, including residential areas, dorms, administration building, cafeteria and the sports ground. Therefore, our ontology has 48 Open-311-Things instances. Open 311 Thing objects were created in Protégé[3] for each request and location permutation. Each Open 311 Thing individual is linked with objects of classes Agency, Subject, Location, 311 types and Action via appropriate object properties, according to the class hierarchy represented by Fig. 1. The ontology in RDF/OWL format was integrated with the OpenLink Virtuoso[4] Universal Server. It is a multi-model data server. Virtuoso was the backend server for the mobile-friendly web application. Agencies have important distinct details linked through data properties such as contact information and governing body. All Open 311 Thing individuals together form a set of all mappings of requests and subjects and locations. In essence, the open 311 Thing individuals encompass all the service requests that are catered by the system. Fig. 2 illustrates the specific instances created in the ontology and their property assertions, including instances of Open311 Thing.

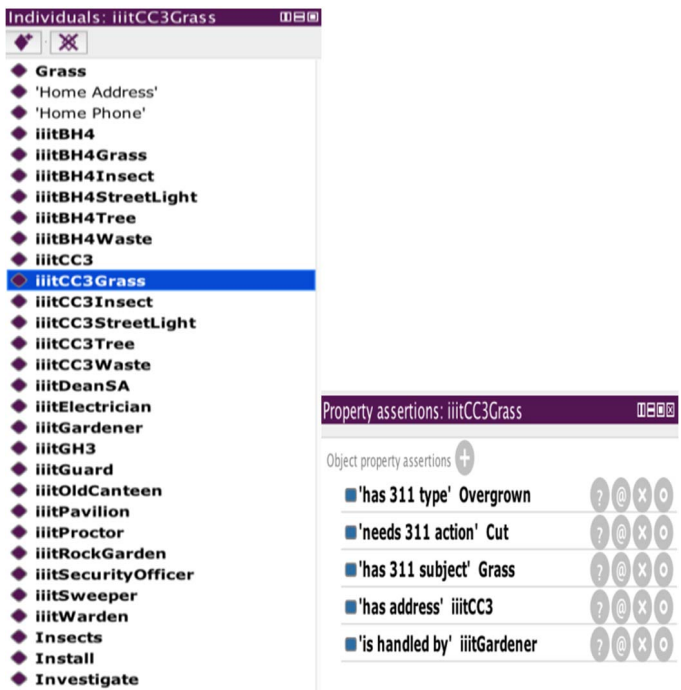

Fig. 2. Individual instances and property assertions.

Python library rdflib[5] connected the application to Virtuoso. The query entered by the citizen was processed by the Natural Language Processing Module of the system. The query was parsed, syntactically and semantically analyzed. It was cleaned of noise through stop word removal and lemmatization. A list of synonyms was prepared which is an exhaustive list of aliases of each subject and location so as to minimize the ambiguity. Local dialect and languages, native to the city, were also considered in the preparation of the list. Keywords were extracted from the query based on the list. Subject and location that were extracted from the query are replaced in appropriate places in a pre-made SPARQL query.

The SPARQL query ran on Virtuoso and returned corresponding results which include the action and agency responsible for the event. The query was stored in the database for data analysis. It included data like unique ID, event, agency, action, status and information about the reporter. An email was also sent to the agency regarding the reported matter. The contact information and other important details were shown to the citizen who could act on his/her behest. A sequential flow of the entire system has been represented in Fig. 3. Thus, a citizen is able to report a non-emergency event.

IV. RESULTS

The evaluation of the ontology is performed in two parts: the first part analyzes the ability of the ontology to represent the data such that the citizen would be able to retrieve his desired information. In the second part, the ontology is evaluated by analyzing how citizen complaints/requests are handled by the ontology. For this, we put the ontology to test by firing SPARQL queries to some common competency questions (CQ).

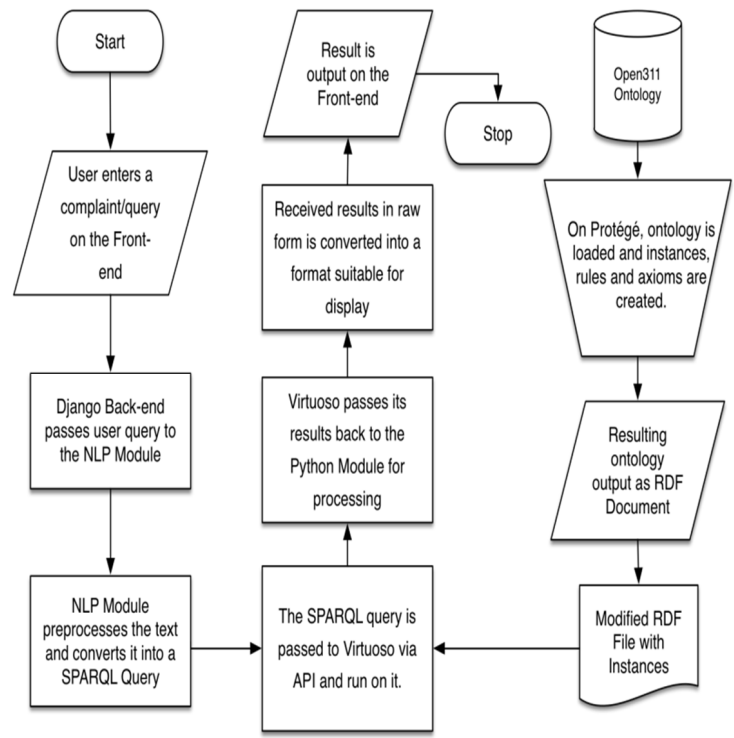

Fig. 3. The proposed workflow of the system

## A. Mapping Citizen Requests to the Open311 Ontology

In this section we illustrate how the citizen queries/requests which are converted to legal SPARQL queries are mapped by our system to the Open311 ontology. Each Open311 Thing has 5 property assertions, which are linked in the manner as represented by Fig. 4. In the example above, we have shown an Open311 Thing 'Overgrown Grass near Computer Center III's. The instances has address, which is 'iiitCC3', 311Type which is 'Overgrown', 311Subject which is 'Grass', needs 311Action 'Cut' and is handled by 'iiitGardener'.

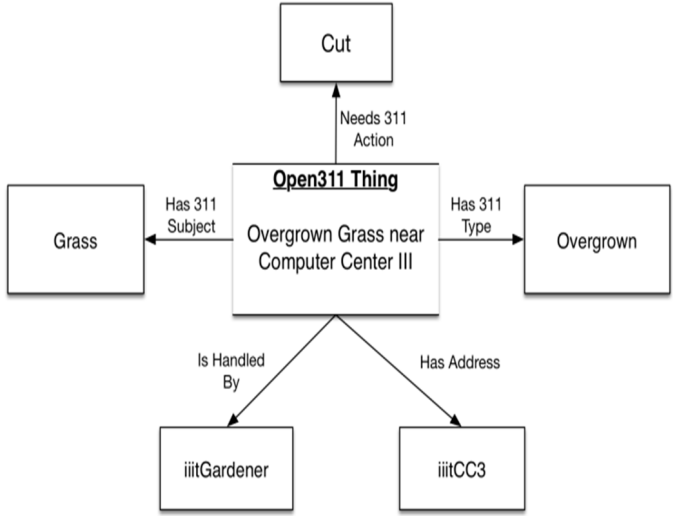

Fig. 4. Mapping of user queries to the Open311 Ontology

## B. Performance Analysis using SPARQL Queries

This section scrutinizes the ontology by testing how the ontology would perform if the SPARQL query language would be used to retrieve relevant data from the ontology and answer the specific questions that we would like it to.

**CQ-1:** Which agency is responsible for a given service request at a given location and what should be the action taken for redressal?

In order to answer the above question, we would require to extract the name of the agency and the action taken, given a service request and a location. The SPARQL query in Fig. 5 performs the task:

```
PREFIX O311O: <http://ontology.eil.utoronto.ca/open311.owl#>
SELECT * WHERE{
    ?subject  O311O:has Address O311O:location.
    ?subject O311O:has311Subject O311O:requestSubject.
    ?subject O311O:isHandledBy ?authority.
    ?subject O311O:need311Action ?action
}
```

Fig. 5. SPARQL Query for CQ-1

In our ontology, the Open311 Thing has properties: location, subject, authority, action and type. The above SPARQL query would return the result given by Fig. 6.

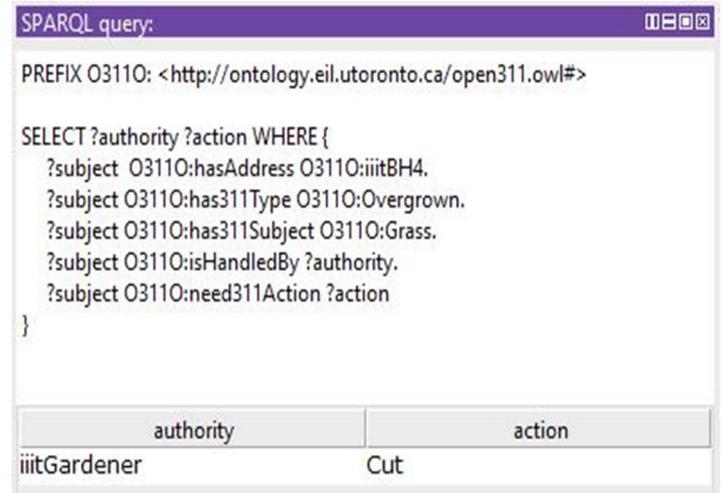

Fig. 6. The result obtained on a fire SPARQL query for CQ-1

**CQ-2:** Find the types of requests and complaints handled by an Authority or Agency.

In order to answer the above question, we would require to extract the name of all the requests and complaints handled by a single agency. In the SPARQL query given in Fig. 7, we retrieve the complaints handled by the IIIT Electricity Department:

```
PREFIX O311O: <http://ontology.eil.utoronto.ca/open311.owl#>
SELECT ?subject ?type WHERE {
    ?thing O311O:has311Subject ?subject.
    ?thing O311O:has311Type ?type.
    ?thing O311O:isHandledBy O311O:iiitElectrician.
}
```

Fig. 7. SPARQL Query for CQ-2

The query in Fig. 7, when fired on the ontology, would return the result illustrated in Fig. 8.

```
SPARQL query:
PREFIX O311O: <http://ontology.eil.utoronto.ca/open311.owl#>

SELECT ?subject ?type WHERE {
   ?thing O311O:has311Subject ?subject.
   ?thing O311O:has311Type ?type.
   ?thing O311O:isHandledBy O311O:iiitElectrician.
}
```

| subject | type |
|---|---|
| 'Street Light' | Damaged |
| 'Street Light' | Damaged |

Fig. 8. The result obtained on a fire SPARQL query for CQ-2

**CQ-3**: Find what authorities or agencies are responsible for a complaint for a particular location.

In order to answer the above question, we would require to extract the name of the authority responsible for handling requests at a particular location. The SPARQL query given in Fig. 9 performs the task:

```
PREFIX O3110: <http://ontology.eil.utoronto.ca/open311.owl#>

SELECT ?agency ?subject WHERE {
   ?thing o311o:hasAddress O311O:location.
   ?thing O311O:isHandledBy ?agency.
   ?thing O311O:has311Subject ?subject
}
```

Fig. 9. SPARQL Query for CQ-3

The query in Fig. 9, when fired on the ontology, would return the result as illustrated in Fig. 10.

```
SPARQL query:
PREFIX O311O: <http://ontology.eil.utoronto.ca/open311.owl#>

SELECT ?agency ?subject WHERE {
   ?thing O311O:hasAddress O311O:iiitCC3.
   ?thing O311O:isHandledBy ?agency.
   ?thing O311O:has311Subject ?subject
}
```

| agency | subject |
|---|---|
| iiitElectrician | 'Street Light' |
| iiitGardener | Grass |
| iiitSweeper | Waste |
| iiitGardener | Tree |
| iiitGuard | Insects |

Fig. 10. The result obtained on a fire SPARQL query for CQ-3

**CQ-4:** Find what are the respective authorities or agencies that are responsible for a handling a request at different locations.

In order to answer the above question, we would require to extract the name of the authorities responsible for handling particular requests at different locations. The SPARQL query in Fig. 11 performs the task:

```
PREFIX O3110: <http://ontology.eil.utoronto.ca/open311.owl#>

SELECT ?agency ?location WHERE {
       ?thing  O3110:hasAddress ?location.
       ?thing O3110:isHandledBy ?agency.
       ?thing O3110:has311Subject O3110:subject
}
```

Fig. 11. SPARQL Query for CQ-4

The query in Fig. 11, when fired on the ontology, would return the result illustrated in Fig. 12:

```
SPARQL query:
PREFIX O311O: <http://ontology.eil.utoronto.ca/open311.owl#>

SELECT ?agency ?location WHERE {
   ?thing  O311O:hasAddress ?location.
   ?thing O311O:isHandledBy ?agency.
   ?thing O311O:has311Subject O311O:Grass.
}
```

| agency | location |
|---|---|
| iiitGardener | iiitBH4 |
| iiitGardener | iiitCC3 |

Fig. 12. The result obtained on a fire SPARQL query for CQ-4

## V. DISCUSSION

Our approach has certain key advantages as follows:

- A semantic analysis of the queries, that allows the system to understand the context of the queries, in contrast with the present day search engines that primarily rank results on string match and association with other pages.
- Schema flexibility that allows the approach to add a variety of subjects with ease thus enabling the approach to handle all the possible non-emergency events that a citizen of a smart city may face.
- Interoperability that facilitates the implementation of the approach to any city in India with just minor changes, to adapt to the possible different subjects. The reuse of this ontology that interoperability permits can cut down significantly the cost of redevelopment for each city.
- Natural language queries and the every citizen can report an event without the technical knowledge of structuring complex queries.

The challenges of our approach are as follows:
- The current implementation models a small domain. The instances and scenario tested during the development are limited to the scope of a university.
- Certain different queries might be identical to the natural language processing module.
- Certain queries may result in incorrect redressal information if the domain of the query shares keywords with another domain. For example the queries, "There is stagnant water in front of the Computer Lab" and "Storm water has the entered the Computer Lab" may have different authorities responsible but the natural language processing module might give out the same result.
- Need for feedback mechanism: A feedback mechanism would help the authorities better serve the community. The mechanism would allow the citizens to interact with the authorities and provide them with the valuable information on the ground realities. The present ontology needs to be modified to support the feedback mechanism.

## VI. CONCLUSION AND FUTURE WORK

In this paper an efficient way to handle a wide range of non-emergency events in a smart city has been introduced and we conclude that Open 311 Ontology, on creating instances representing the city is able to classify the non-emergency events in an organized manner, providing a semantic model of the city. Based on the preliminary results, the ontology can be systematically queried by the citizen with queries that essentially characterize the day-to-day events that are faced by the citizen in a city, to obtain effective resolutions. The essence of the presented approach is the fact that a wide range of queries in natural language of a typical user, can be analyzed and processed to query the semantic model of assorted and distinct possible events, ultimately providing a single endpoint to the citizen.

The approach has an immense scope of future work, being in its infancy, major additions could be made to make the approach more robust in handling the diverse queries that a citizen residing in a smart city may have. A possible direction for future work is the implementation of a consolidated view of queries that would provide decision making information to the authorities. Another direction would be extending the ontology to support feedback from the citizens.

ACKNOWLEDGMENT

The authors would like to acknowledge the support and guidance of Prof. Nathalie Mitton (Scientific Head of the FUN research group at INRIA Lille-Nord Europe). The discussions we had with her were significantly helpful in our research. We are indebted to her contribution. We would also like to thank Riccardo Petrolo of INRIA Lille-Nord Europe for his invaluable feedback and guidance.